%% file: robustConstitutiveModelCalib_21Feb23.tex
\RequirePackage{tikz} 

\documentclass[sn-mathphys]{sn-jnl}

\jyear{2023}%

\theoremstyle{thmstyleone}%
\theoremstyle{thmstyletwo}%

\theoremstyle{thmstylethree}%

\usepackage[capitalise, noabbrev, nameinlink]{cleveref}
\usepackage{listings}
\usepackage{epsfig}
\usepackage{amsmath}
\usepackage{amsfonts}
\usepackage{amsbsy}
\usepackage{amsthm}
\usepackage{amssymb}
\usepackage{subcaption}
\usepackage{lipsum}
\usepackage[utf8]{inputenc}
\usepackage[T1]{fontenc}
\usepackage{enumerate}
\usepackage{hyphenat}
\usepackage{bbm}
\usepackage{fdsymbol}
\usepackage{subcaption}
\usepackage{graphicx}
\usepackage{grffile}
\usepackage{algorithm,algorithmicx}
\usepackage{algpseudocode}
\algnewcommand\algorithmicinput{\textbf{Input: }}
\algnewcommand\algorithmicoutput{\textbf{Output: }}
\usepackage{algpseudocode}
    \algnewcommand\algorithmicforeach{\textbf{for each}}
    \algdef{S}[FOR]{ForEach}[1]{\algorithmicforeach\ #1\ \algorithmicdo}
    \algnewcommand\algorithmictimes{\textbf{times}}
    \algdef{E}[REPEAT]{Times}[1]{#1\ \algorithmictimes}
    \algrenewcommand\textproc{\textsf}
    \algnewcommand{\IfThen}[2]{\State \algorithmicif\ #1\ \algorithmicthen\ #2}
\usepackage{afterpage}
\usepackage{epstopdf}
\usepackage{lineno}
\usepackage[normalem]{ulem}

\modulolinenumbers[5]
\usepackage{mathtools}

\usepackage{hyperref}
\hypersetup{colorlinks=true}

\everymath{\displaystyle}

\DeclareMathOperator*{\argmax}{argmax}

\newcommand\E[1]{{\mathbb{E}\left[#1\right]}}

\usepackage{program} 
\usepackage{smartdiagram} 
\usesmartdiagramlibrary{additions}
\usetikzlibrary{shapes.geometric,arrows,quotes}

\newcommand{\lpipe}{\rule[-0.4ex]{0.41pt}{2.3ex}} 
\newcommand{\tlpipe}{\rule[-1.8ex]{0.41pt}{5.0ex}} 

\begin{document}


\title[Asynchronous parallel constitutive model calibration for CPFEM]{An asynchronous parallel high-throughput model calibration framework for crystal plasticity finite element constitutive models}


\author*[1]{\fnm{Anh} \sur{Tran}}\email{anhtran@sandia.gov}


\author[2]{\fnm{Hojun} \sur{Lim}}\email{hnlim@sandia.gov}

\affil*[1]{\orgdiv{Scientific Machine Learning}, \orgname{Sandia National Laboratories}, \orgaddress{\city{Albuquerque}, \state{NM}, \postcode{87123}, \country{USA}}}


\affil[2]{\orgdiv{Computational Materials \& Data Science}, \orgname{Sandia National Laboratories}, \orgaddress{\city{Albuquerque}, \state{NM}, \postcode{87123}, \country{USA}}}


\abstract{
\textcolor{black}{
Crystal plasticity finite element model (CPFEM) is a powerful numerical simulation in the integrated computational materials engineering (ICME) toolboxes that relates microstructures to homogenized materials properties and establishes the structure-property linkages in computational materials science.
However, to establish the predictive capability, one needs to calibrate the underlying constitutive model, verify the solution and validate the model prediction against experimental data. 
Bayesian optimization (BO) has stood out as a gradient-free efficient global optimization algorithm that is capable of calibrating constitutive models for CPFEM. 
In this paper, we apply a recently developed asynchronous parallel constrained BO algorithm to calibrate phenomenological constitutive models for stainless steel 304L, Tantalum, and Cantor high-entropy alloy.
}
}




\maketitle

\section{Introduction}
\label{sec:Introduction}

Researching and developing predictive integrated computational materials engineering (ICME) models at multiple length-scales and time-scales has been conducted at the last two decades. 
In order to establish the predictive capability, the ICME model must be numerically verified and experimentally validated, in the spirit of uncertainty quantification. 
The development of ICME models as the third paradigm and scientific machine learning as the fourth paradigm~\cite{hey2009fourth,agrawal2016perspective} is consistent with the Materials Genome Initiative~\cite{national2011materials,holdren2014materials,lander2021materials}, which has long been held as a cornerstone in materials science. 

In the ICME context, crystal plasticity finite element model (CPFEM), along with phase-field simulations, have been the main workhorses of computational materials science to investigate the structure-property relationship. 
By targeting the mesoscale, which is sandwiched between microscale and macroscale, they are also important toolboxes in tackling multiscale ICME models. 
Numerous constitutive models have been proposed over the \textcolor{black}{past few decades, including but not limited to}~\cite{roters2019damask}, isotropic $J_2$ plasticity, phenomenological crystal plasticity, dislocation-density-based crystal plasticity, atomistically-informed crystal plasticity, crystal plasticity including dislocation flux. 
Depending on the materials system of interest and its corresponding crystal structures, as well as the parameterization of the constitutive model used, there are many parameters underpinning the constitutive model that one may need to calibrate against a suitable experimental setup. 

Constitutive model calibration has been fairly well-studied within the literature. 
Loosely speaking, calibrating a constitutive model for CPFEM can be formulated as an optimization problem, which in turn minimizes the misfit between the CPFEM and experimental data. 
However, \textcolor{black}{most approaches}, if not all as of this point, mainly utilized sequential optimization, which suffers from the high computational cost of the forward CPFEM model. 
Even though the high computational cost can be somewhat mitigated by message passing interface (MPI) parallelism on multi-core high-performance computers, it also follows a diminishing return characterized by Amdahl's law~\cite{hill2008amdahl}. 
Modern optimization approaches often exploit both MPI (or MPI+X for heterogeneous) parallelism as well as optimization parallelism on high-performance computers and are capable of delivering a better solution in shorter wallclock time. 
Compared to sequential optimization approaches, parallel optimization approaches can offer better performance in calibrating ICME models, especially for prototyping models that do not scale well with multi-cores and multi-threads on high-performance computing platforms. 

Since the birth of CPFEM~\cite{raabe1998computational,raabe2004continuum,janssens2010computational,roters2011crystal}, there have been numerous works in the literature dedicated to calibrating constitutive model in many materials system. 
Due to limited space, we attempt to review some notable works that are relevant to constitutive model calibration in the context of CPFEM. 
Chakraborty and Eisenlohr~\cite{chakraborty2017evaluation} proposed a modified Nelder--Mead optimization algorithm to treat the objective function when evaluation fails and demonstrated their algorithm on the single crystal nano-indentation for face-centered cubic materials systems. 
H{\'e}rault et al.~\cite{herault2021calibration} applied the Levenberg-Marquadt gradient-based optimization algorithm to calibrate an isotropic constitutive model for dual phase steel DP600. 
Nguyen et al.~\cite{nguyen2021bayesian} applied Markov chain Monte Carlo on top of a Gaussian process (GP) regression to obtain a posterior of a dislocation-density-based constitutive model parameters for copper. 
Savage et al.~\cite{savage2021identification} applied genetic algorithm multi-objective optimization algorithm to calibrate a dislocation-density-based constitutive model for dual phase 780 steel against multiple criteria. 
Hochhalter et al.~\cite{hochhalter2020non} applied the Nelder--Mead and Broyden--Fletcher--Goldfarb--Shanno optimization algorithms before running Markov Chain Monte Carlo to obtain the posterior of a phenomenological constitutive model parameters for aluminum oligocrystal. 
Kuhn et al.~\cite{kuhn2022identifying} also utilized Bayesian optimization (BO) to calibrate a phenomenological constitutive model for 50CrMo4 steel. 
Sedighiani et al.~\cite{sedighiani2020efficient,sedighiani2022determination} applied genetic algorithm to calibrate both phenomenological and dislocation-density-based constitutive models for face-centered cubic copper, body-centered cubic steel, and hexagonal-close packed magnesium. 
Wang et al.~\cite{wang2016identifying} employed the gradient-based least-squares optimization algorithm to calibrate a micro-polar plasticity in Hostun sand. 
Liu et al.~\cite{liu2016determining} employed a Gauss--Newton trust region optimization algorithm to calibrate a phenomenological Dafalias--Manzari constitutive model for Nevada sand. 
Herrera-Solaz et al.~\cite{herrera2014inverse} used the Levenberg--Marquadt optimization algorithm to calibrate a phenomenological constitutive model for AZ31 Magnesium alloy. 
Do and Ohsaki~\cite{do2022bayesian,do2022proximal} proposed a multi-objective BO to calibrate an elastoplastic constitutive model for a structural steel. 
Seidl and Granzow~\cite{seidl2022calibration} proposed an automatic differentiation framework to calibrate parameters in elastoplastic constitutive model. 
Corona et al.~\cite{corona2021anisotropic} calibrated three yield functions, von Mises, Hill-48, and Yld2004-18p on a 7079 aluminum alloy against digital image correlation experimental data, which were subsequently verified and validated in Jones et al~\cite{jones2021anisotropic}.
Karandikar et al.~\cite{karandikar2022bayesian} also applied BO to calibrate a Johnson-Cook flow stress model. 
Sun and Wang~\cite{sun2022method} applied BO to calibrate a Voce hardening law for Magnesium alloy ZEK100 in a viscoplastic self-consistent polycrystal plasticity model with twinning and de-twinning scheme. 
Morand and Helm~\cite{morand2019mixture} utilized a mixture of experts with an ensemble of neural networks to calibrate an exponential hardening model. 
Generale et al.~\cite{generale2022bayesian} applied parallel MCMC to calibrate a viscous multimode damage model in an oxide-oxide ceramic matrix composite. 
Zambaldi et al.~\cite{zambaldi2012orientation} applied the Nelder--Mead optimization algorithm to calibrate a phenomenological constitutive model for $\alpha$ Titanium. 
Bolzon et al.~\cite{bolzon2004material} applied the trust region optimization algorithm to calibrate an elastoplastic constitutive model for ductile metals. 
Fuhg et al. utilized the partially convex neural network~\cite{fuhg2022machine} and local GP~\cite{fuhg2022local} to derive a data-driven macroscopic yield function. 
Zhang et al.~\cite{zhang2022method} calibrated a Johnson-Cook constitutive model in advanced high strength steel DP1180 against nano-indentation experimental data. 
\textcolor{black}{
Foumani et al.~\cite{foumani2022multi} proposed a multi-fidelity BO approach based on latent map GP and applied on nanolaminate ternary alloy family as well as organic-inorganic perovskite datasets. 
Wang et al.~\cite{wang2020parallel} proposed a q-EI algorithm to parallelize BO in an asynchronous manner and analyzed the convergence accordingly. 
Veasna et al.~\cite{veasna2023machine} proposed to couple SMS-EGO~\cite{ponweiser2008multiobjective}, which is a hypervolume-based approach, to calibrate constitutive model for elasto-plastic self-consistent CPFEM model. 
Bostanabad et al.~\cite{bostanabad2018leveraging} incorporated and optimized the nugget parameter as an additional hyperparameter for noisy environment, such as experimental observations. 
}

Despite the fact that numerous constitutive model calibration works have been done in the literature, not too many have taken a parallel optimization route that offers greater efficiency in terms of computational cost. 
In this paper, we apply a recently developed asynchronous parallel constrained BO algorithm~\cite{tran2022aphbo}, built upon the GP-Hedge algorithm~\cite{brochu2010portfolio}, to calibrate \textcolor{black}{phenomenological constitutive models} for stainless steel 304L, \textcolor{black}{Tantalum, and Cantor high-entropy alloy}. Compared to other previous works listed above, one of the significant advantages of this framework is the asynchronous parallel feature, which allows multiple CPFEM simulations running asynchronous parallel on high-performance computers to optimize the objective in shorter wallclock time. 

The remainder of the this paper is organized as follows. 
Section~\ref{sec:BO} introduces the BO algorithm used in this paper, its underlying GP and the adaptive sampling strategy for finding the next query point. 
Section~\ref{sec:CPFEM} describes the phenomenological constitutive model, based on the open-source \texttt{DAMASK} CPFEM package~\cite{roters2019damask}. 
Section~\ref{sec:ModelCalibrationWorkflow} introduces the asynchronous parallel constitutive model calibration workflow used in this paper. 
Section~\ref{sec:SS304L} describes the numerical and experimental results for stainless steel 304L. 
Section~\ref{sec:Tantalum} describes the numerical and experimental results for Tantalum. 
Section~\ref{sec:CantorHEA} describes the numerical and experimental results for Cantor high-entropy alloy. 
Section~\ref{sec:DiscussionConclusion} discusses and concludes the paper.




\section{Bayesian optimization}
\label{sec:BO}

In this section, we introduce the BO concept, which is based on GP and an asynchronous parallel sampling strategy that allows multiple acquisition functions simultaneously. 

\subsection{Gaussian process regression and Bayesian optimization}

Comprehensive and critical review studies are provided by Brochu et al.~\cite{brochu2010tutorial}, Shahriari et al.~\cite{shahriari2016taking}, Frazier~\cite{frazier2018tutorial}, and Jones et al.~\cite{jones1998efficient} for BO method and its variants. 
We adopt the notation from Shahriari et al.~\cite{shahriari2016taking} and Tran et al.~\cite{tran2022aphbo,tran2020smfbo2cogp,tran2020weargp,tran2019sbfbo2cogp,tran2019pbo,tran2019constrained,tran2018efficient,tran2018weargp}. 
In this formulation, we treat the optimization problem in the maximization setting,
\begin{equation}
\mathbf{x}^\star = \argmax_{x\in\mathcal{X}} f(\mathbf{x})
\end{equation}
subject to a set of nonlinear inequalities constraints
$\lambda_j (\mathbf{x}) \leq 0, \quad 1 \leq j \leq J$, assuming that $f$ is a function of $\mathbf{x}$, where $\mathbf{x} \in \mathcal{X} \subset \mathbb{R}^d$ is the $d$-dimensional input. A $\mathcal{GP}(\mu_0,k)$ is a non-parametric model over functions $f$,
which is fully characterized by the prior mean functions $\mu_0(x): \mathcal{X} \mapsto \mathbb{R}$ and the positive-definite kernel, or covariance function $k:\mathcal{X} \times \mathcal{X} \mapsto \mathbb{R}$. 

Let the dataset $\mathcal{D}={(\mathbf{x}_i,y_i)}_{i=1}^n$ denote a collection of $n$ noisy observations. 
In GP regression, it is assumed that $\mathbf{f} = f_{1:n}$ is jointly Gaussian, and the observation $y$ is normally distributed given $f$, leading to  
\begin{equation}
\label{eq:prior}
\mathbf{f} \lpipe \mathbf{x} \sim \mathcal{N}(\mathbf{m},\mathbf{K}),
\end{equation}
\begin{equation}
\mathbf{y} \lpipe \mathbf{f},\sigma^2 \sim \mathcal{N}(\mathbf{f},\sigma^2 \mathbf{I}),
\end{equation}
where $m_i := \mu(\mathbf{x}_i)$, and $K_{i,j} := k(\mathbf{x}_i,\mathbf{x}_j)$. Equation~\ref{eq:prior} describes the prior distribution induced by the GP.

The covariance kernel $k(\cdot, \cdot)$ is a choice of modeling covariance between inputs, depending on the smoothness assumption of $f$. The family of Mat{\'e}rn kernels is arguably one of the most popular choices for kernels, offering a broad class for stationary kernels which are controlled by a smoothness parameter $\nu>0$ (cf. Section 4.2,~\cite{rasmussen2006gaussian}), including the square-exponential ($\nu \to \infty$) and exponential $(\nu = 1/2)$ kernels. The Mat{\'e}rn kernels are described as
\begin{equation}
\mathbf{K}_{i,j} = k(\mathbf{x}_i, \mathbf{x}_j) = \theta_0^2 \frac{2^{1-\nu}}{\Gamma(\nu)} (\sqrt{2\nu} r)^{\nu} K_{\nu}(\sqrt{2\nu}r), 
\end{equation}
where $K_\nu$ is a modified Bessel fuction of the second kind and order $\nu$. 
Common kernels for GP include~\cite{shahriari2016taking}
\begin{itemize}
\item $\nu = 1/2: k_{\text{Mat{\'e}rn}1/2} (\mathbf{x}, \mathbf{x'}) = \theta_0^2 \exp{(-r)}$ (also known as exponential kernel), 
\item $\nu = 3/2: k_{\text{Mat{\'e}rn}3/2} (\mathbf{x}, \mathbf{x'}) = \theta_0^2 \exp{(-\sqrt{3}r)} (1+\sqrt{3} r)$, 
\item $\nu = 5/2: k_{\text{Mat{\'e}rn}5/2} (\mathbf{x}, \mathbf{x'}) = \theta_0^2 \exp{(-\sqrt{5}r)} \left( 1 + \sqrt{5}r + \frac{5}{3}r^2 \right)$, 
\item $\nu \to \infty: k_{\text{sq-exp}} (\mathbf{x}, \mathbf{x'}) = \theta_0^2 \exp{\left(-\frac{r^2}{2} \right)}$ (also known as square exponential or automatic relevance determination kernel),
\end{itemize}
where $r^2 = (\mathbf{x} - \mathbf{x}')\mathbf{\Lambda}(\mathbf{x} - \mathbf{x}')$, and $\mathbf{\Lambda}$ is a diagonal matrix of $d$ squared length scale $\theta_i$. 


Under the formulation of GP, given the dataset $\mathcal{D}_n$, the prediction for an unknown arbitrary point is characterized by the posterior Gaussian distribution, which can be described by the posterior mean and posterior variance functions, respectively as 
\begin{equation}
\label{eq:posteriorMean}
\mu_n(\mathbf{x}) = \mu_0(\mathbf{x}) + \mathbf{k}(\mathbf{x})^T (\mathbf{K} + \sigma^2 \mathbf{I})^{-1} (\mathbf{y} - \mathbf{m}),
\end{equation}
and
\begin{equation}
\label{eq:posteriorVariance}
\sigma_n^2(\mathbf{x}) = k(\mathbf{x}, \mathbf{x}) - \mathbf{k}(\mathbf{x})^T (\mathbf{K}  + \sigma^2 \mathbf{I})^{-1} \mathbf{k}(\mathbf{x}),
\end{equation}
where $\mathbf{k}(\mathbf{x})$ is the covariance vector between the query point $\mathbf{x}$ and $\mathbf{x}_{1:n}$. 
The main drawback of GP formulation is its scalability $\mathcal{O}(n^3)$ that originates from the computation of the inverse of the covariance matrix $\mathbf{K}$. 
Connection from GP to convolution neural network has been proposed where it is proved to be theoretically equivalent to single layer with infinite width~\cite{lee2017deep} or infinite convolutional filters~\cite{garriga2018deep}.

\subsection{Acquisition function}
\label{subsec:acquisitionFunction}

In the traditional BO method, which is sequential, the GP model is constructed for the objective function, and the next sampling location is determined by maximizing the acquisition function based on the constructed GP. 
This acquisition function is evaluated based on the underlying GP surrogate model, thus converting the cost of evaluating the real simulation to the cost of evaluating on the GP model. 
The \textcolor{black}{latter} is much more computationally appealing because it is semi-analytical. 
The acquisition function must balance between the exploitation and exploration flavors of the BO method. 
Too much exploitation would drive the numerical solution to a local minima, whereas too much exploration would make BO an inefficient optimization method. 
We review three main acquisition functions that are typically used in the literature: the probability of improvement (PI), the expected improvement (EI), and the upper confidence bounds (UCB). 

Denote $\mu(\mathbf{x})$, $\sigma^2(\mathbf{x})$, and $\theta$ as the posterior mean, the posterior variance, and the hyper-parameters of the objective GP model, respectively. 
$\theta$ is obtained by maximizing the log likelihood estimation over a plausible chosen range, where the likelihood is described as
\begin{equation}
\begin{array}{lll}
\log{p(\mathbf{y} \lpipe \mathbf{x}_{1:n}, \theta )} &=& - \frac{n}{2} \log{(2\pi)} \\
&& - \frac{1}{2} \log{\lpipe \mathbf{K}^{\theta} + \sigma^2 \mathbf{I} \lpipe} \\
&& - \frac{1}{2} (\mathbf{y} - \mathbf{m}_{\theta})^T (\mathbf{K}^{\theta} + \sigma^2 \mathbf{I} )^{-1} (\mathbf{y} - \mathbf{m}_{\theta}).
\end{array}
\end{equation}
Let $\phi(\cdot)$ and $\Phi(\cdot)$ be the standard normal probability distribution function and cumulative distribution function, respectively, and $\mathbf{x}_{\text{best}} = \argmax_{1\leq i \leq n} f(\mathbf{x}_i)$ be the best-so-far sample. 
Rigorously, the acquisition function should be written as $\alpha (\mathbf{x};\{\mathbf{x}_i,y_i \}_{i=1}^n,\theta)$, but for the sake of simplicity, we drop the dependence on the observations and simply write as $\alpha(\mathbf{x})$ and $\mathbb{E}(\cdot)$ is implicitly understood as $\mathbb{E}_{y \sim p(y \lpipe \mathcal{D}_n, \mathbf{x})} (\cdot)$ unless specified otherwise.

The PI acquisition function~\cite{kushner1964new} is defined as 
  \begin{equation}
  \alpha_{\text{PI}}(\mathbf{x}) = \text{Pr}(y > f(\mathbf{x}_\text{best})) = \E{\mathbbm{1}_{y>f(\mathbf{x}_\text{best})}} = \Phi(\gamma(\mathbf{x})),
  \end{equation}
where
  \begin{equation}
  \label{eq:normalizedZ}
  \gamma(\mathbf{x}) = \frac{\mu(\mathbf{x}) - f(\mathbf{x}_{\text{best}})}{\sigma(\mathbf{x})},
  \end{equation}
indicates the deviation away from the best sample. 
The PI acquisition function is constructed based on the idea of binary utility function, where a unit reward is received if a new best-so-far sample is found, and zero otherwise. 

The EI acquisition function~\cite{mockus1975bayesian,mockus1982bayesian,jones2001taxonomy,bull2011convergence,snoek2012practical} is defined as 
\begin{equation}
\alpha_{\text{EI}}(\mathbf{x}) = \sigma(\mathbf{x}) \left[ \gamma(\mathbf{x}) \Phi(\gamma(\mathbf{x})) + \phi(\gamma(\mathbf{x})) \right].
\end{equation}
The EI acquisition is constructed based on an improvement utility function, where the reward is the relative difference if a new best-so-far sample is found, and zero otherwise. 
A closely related generalization of the EI acquisition function, called knowledge-gradient (KG) acquisition function, has been suggested in~\cite{scott2011correlated}. Under the assumptions of noise-free and the sampling function is restricted, the EI acquisition function is recovered from the KG acquisition function. 
If the EI acquisition function is rewritten as 
\begin{equation}
\begin{array}{lll}
\alpha_\text{EI}(\mathbf{x}) &=& \E{\max\left(y, f(\mathbf{x}_\text{best})\right) - f(\mathbf{x}_\text{best})} \\
&=& \E{\max(y - f(\mathbf{x}_\text{best}), 0} \\ 
&=& \E{(y - f(\mathbf{x}_\text{best})^+},
\end{array}
\end{equation} 
then the KG acquisition function is expressed as 
\begin{equation}
\alpha_\text{KG}(\mathbf{x}) = \E{ \max \mu_{n+1}(\mathbf{x}) \lpipe \mathbf{x}_{n+1} = \mathbf{x}} - \max(\mu_n(\mathbf{x}))
\end{equation}
for one-step look-ahead acquisition function. 

The UCB acquisition function~\cite{auer2002using,srinivas2009gaussian,srinivas2012information} is defined as
\begin{equation}
\alpha_{\text{UCB}}(\mathbf{x}) = \mu(\mathbf{x}) + \kappa \sigma(\mathbf{x}),
\end{equation}
where $\kappa$ is a hyper-parameter describing the acquisition exploitation-exploration balance. 
Here, we adopt the $\kappa$ computation from Daniel et al.~\cite{daniel2014active}, where
\begin{equation}
\kappa = \sqrt{\nu \gamma_n},\quad \nu = 1, \quad \gamma_n = 2\log{\left(\frac{n^{d/2 + 2}\pi^2}{3\delta} \right)},
\end{equation}
and $d$ is the dimensionality of the problem, and $\delta \in (0,1)$~\cite{srinivas2012information}. 

Another type acquisition function is entropy-based, such as GP-PES~\cite{hernandez2014predictive,hernandez2015predictive,hernandez2016predictive,hernandez2016general}, GP-ES~\cite{hennig2012entropy}, GP-MES~\cite{wang2017max}. Since the GP is collectively a distribution of functions, the distribution of the global optimum $\mathbf{x}^\star$ can be estimated as well from sampling the GP posterior. The predictive-entropy-search (PES)~\cite{hernandez2014predictive} can be written in terms of the differential entropy $H(\cdot)$ as
\begin{equation}
\begin{array}{lll}
\alpha_\text{PES}(\mathbf{x}) &=& H[p(\mathbf{x}^\star \lpipe \mathcal{D}_n)] - \mathbb{E} \left[ H [p(\mathbf{x}^\star \lpipe \mathcal{D} \cup \{ (\mathbf{x},y) \}) ]  \right] \\
&=& H[p(y \lpipe \mathcal{D}_n, \mathbf{x})] - \mathbb{E}_{p(\mathbf{x}^\star \lpipe \mathcal{D}_n)} [H(p(y \lpipe \mathcal{D}_n, \mathbf{x}, \mathbf{x}^\star))] 
\end{array}
\end{equation}
Wang et al.~\cite{wang2017max} proposed GP-MES acquisition function, which effectively suggests sampling $y^\star$ from the 1-dimensional output space $\mathbb{R}$ using Gumbel distribution instead of $\mathbf{x}$ in the high-dimensional input space $\mathcal{X}$, i.e.
\begin{equation}
\alpha_\text{MES}(\mathbf{x}) = H[p(y \lpipe \mathcal{D}_n, \mathbf{x})] - \mathbb{E} [H(p(y \lpipe \mathcal{D}_n, \mathbf{x}, y^\star))],
\end{equation}

Last but not least, since $f(\mathbf{x})$ is Gaussian, the differential entropy can be simplified to a function of posterior variance $\sigma^2(\mathbf{x})$, i.e.
\begin{equation}
H[p(y \lpipe \mathcal{D}_n, \mathbf{x})] = 0.5 \log\left[ 2\pi e (\sigma^2(\mathbf{x}) + \sigma^2) \right],
\end{equation} 
and thus pure exploration search which targets $\mathbf{x}$ with maximum posterior variance $\sigma^2 (\mathbf{x})$, i.e. $\alpha_\text{PE}(\mathbf{x}) = \sigma^2(\mathbf{x})$ is theoretically identical with the maximum differential entropy~\cite{hernandez2014predictive}. 
Interestingly, Wilson et al.~\cite{wilson2018maximizing} proposed a reparameterization of most commonly used acquisition functions in terms of deep learning convolution kernels. 
Wilson \cite{wilson2018maximizing} presented a reparameterization scheme for different acquisition functions in term of activation function for neural networks, summarizing in the following table. 
\begin{table}[]
\centering
\caption{$1^{+/-}$: right-/left-continuous Heaviside step function; ReLU + sigmoid ($\sigma$) + softmax: activation function; $\Sigma = \mathbf{L} \mathbf{L}^\top $: Cholesky factorization; $\gamma \sim \mathcal{N}(\mathbf{0},\Sigma)$ residual}. 
\begin{tabular}{lll}
\hline
\textbf{Abbr.} & \textbf{Acquisition function $\mathcal{L}$} & \textbf{Reparameterization} \\ \hline
EI  & $\mathbb{E}_\mathbf{y} [\max(\text{ReLU}(\mathbf{y} - \alpha))]$ & $\mathbb{E}_\mathbf{z}[\max(\text{ReLU}(\mathbf{\mu} + \mathbf{L} \mathbf{z} - \alpha))]$ \\
PI  & $\mathbb{E}_\mathbf{y}[\max(1^-(\mathbf{y} - \alpha))]$ &  $\mathbb{E}_\mathbf{z}[\max(\sigma ( \frac{\mathbf{\mu} + \mathbf{L} \mathbf{z} - \alpha}{\tau} ))]$  \\
SR  & $\mathbb{E}_\mathbf{y}[\max (\mathbf{y})]$ & $\mathbb{E}_\mathbf{z}[\max(\mathbf{\mu} + \mathbf{L} \mathbf{z})]$ \\
UCB & $\mathbb{E}_\mathbf{y}[\max (\mathbf{\mu} + \sqrt{\frac{\beta \pi}{2}}  \lpipe \gamma \lpipe )]$ & $\mathbb{E}_\mathbf{z}[ \max (\mathbf{\mu} + \sqrt{\frac{\beta \pi}{2}}  \lpipe \mathbf{L} \mathbf{z} \lpipe )]$ \\
ES  & $-\mathbb{E}_{\mathbf{y}_a}[\text{H}(\mathbb{E}_{\mathbf{y}_b  \lpipe  \mathbf{y}_a} [1^+ (\mathbf{y}_b - \max(\mathbf{y}_b))] )]$ & $-\mathbb{E}_{\mathbf{z}_a}[\text{H}(\mathbb{E}_{\mathbf{z}_b} [\text{softmax} ( \frac{\mathbf{\mu}_{b \lpipe a} + \mathbf{L}_{b \lpipe a} \mathbf{z}_b }{\tau} )] )] $
 \\
KG  & $\mathbb{E}_{\mathbf{y}_a} [\max(\mathbf{\mu}_b + \Sigma_{b,a} \Sigma_{a,a}^{-1} (\mathbf{y}_a - \mathbf{\mu}_a))] $ & $\mathbb{E}_{\mathbf{z}_a} [\max(\mathbf{\mu}_b + \Sigma_{b,a} \Sigma_{a,a}^{-1} \mathbf{L}_a \mathbf{z}_a)] $ \\ \hline
\end{tabular}
\end{table}

\subsection{Constraints}

A review and comparison study is performed by Parr et al.~\cite{parr2012infill} for different schemes to handle constraints using both synthetic and real-world applications. 
Many previous works discussed below in the literature prefer to couple constraint satisfaction problems with the EI acquisition due to its consistent numerical performance. 
Even though constraints can be classified to many types, for example~\cite{digabel2015taxonomy}, known versus unknown, quantifiable versus nonquantifiable, a priori versus simulation, relaxable versus unrelaxable, in this work, we mainly focus on two types of constraints, namely known and unknown. 
Known constraints, which are typically physics-based, do not require to invoke the functional evaluation and are often computationally cheap. 
We model known constraints by penalizing the acquisition function with an indicator function $\mathbb{I}(\mathbf{x})$, depending on whether constraints are violated. 
\begin{equation}
\alpha^{\text{known}}_{\text{constrained}}(\mathbf{x}) = \alpha(\mathbf{x}) \mathbb{I}_{\text{known}}(\mathbf{x}),
\end{equation}
where 
\begin{equation}
\mathbb{I}_{\text{known}}(\mathbf{x}) = 
\begin{cases} 
1, \quad{\lambda(\mathbf{x}}) \leq \mathbf{c} \\
0. \quad{\lambda(\mathbf{x}}) \not \leq \mathbf{c} \\
\end{cases}
\end{equation}
Unknown or hidden constraints usually occurs when outputs of a forward computational model is \textit{NaN} or \textit{Inf}. 
For such cases, we augment the probability of failure by a binary probabilistic classifier into the acquisition function in a soft penalization manner as
\begin{equation}
\alpha^{\text{unknown}}_{\text{constrained}}(\mathbf{x}) = 
\begin{cases}
\alpha(\mathbf{x}), \quad \text{ with } \textrm{Pr}(\textrm{clf}(\mathbf{x}) = 1), \\
0, \quad \text{ with } \textrm{Pr}( \textrm{clf}(\mathbf{x}) = 0). \\
\end{cases}
\end{equation}
The expected acquisition function, therefore, follows as
\begin{equation}
\E{\alpha^{\text{unknown}}_{\text{constrained}}(\mathbf{x})} = \alpha(\mathbf{x}) \textrm{Pr}_{\text{unknown}}(\textrm{clf}(\mathbf{x}) = 1). 
\end{equation}

\subsection{Asynchronous parallel}

\begin{figure}[!htbp]
\centering
\subcaptionbox{Batch-sequential parallel BO~\cite{tran2019pbo}.
}
  [.45\linewidth]{\includegraphics[width=0.475\textwidth, keepaspectratio]{
  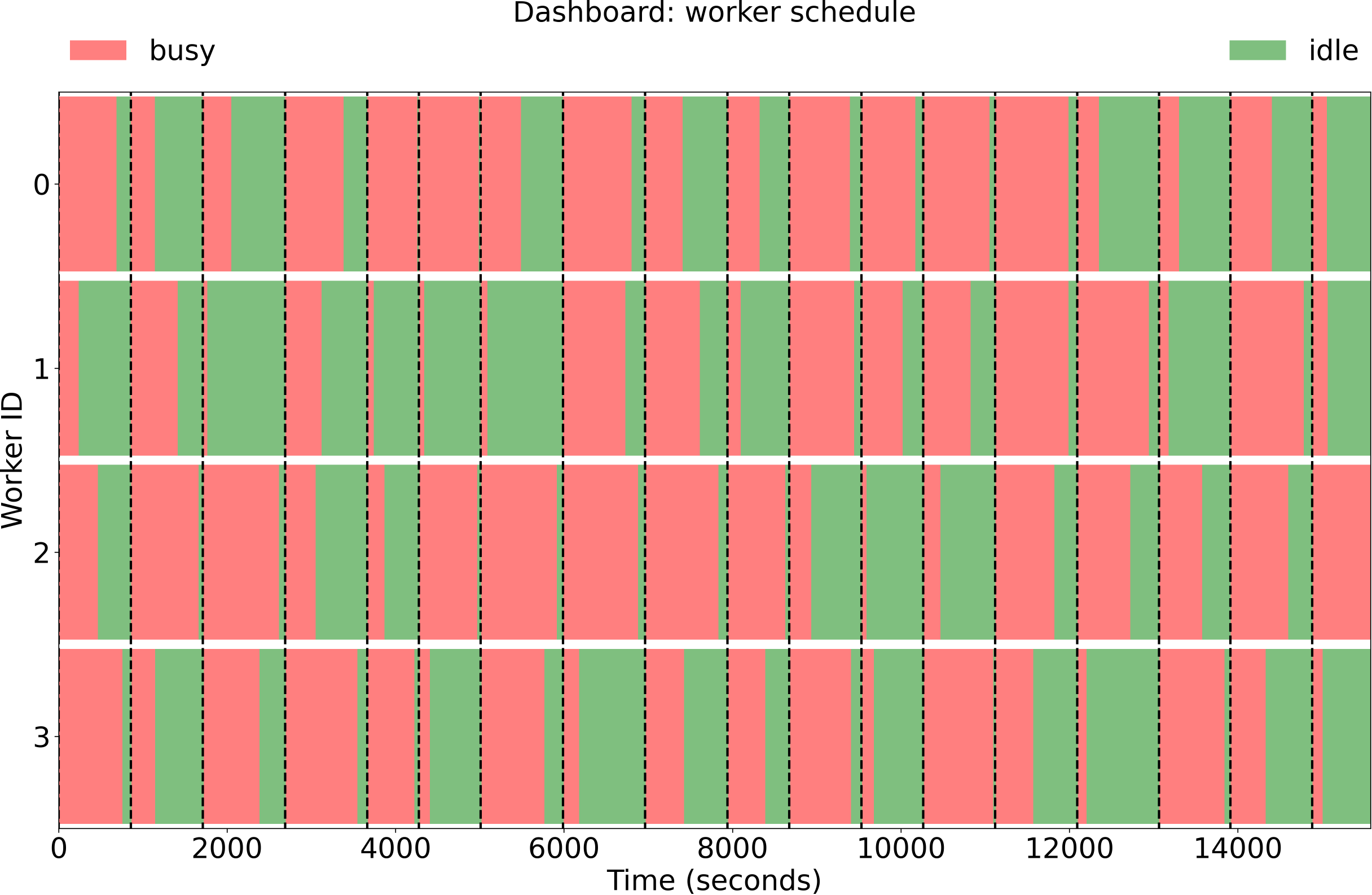}}
\hfill
\subcaptionbox{Asynchronous parallel BO~\cite{tran2022aphbo}.
}
  [.45\linewidth]{\includegraphics[width=0.475\textwidth, keepaspectratio]{
  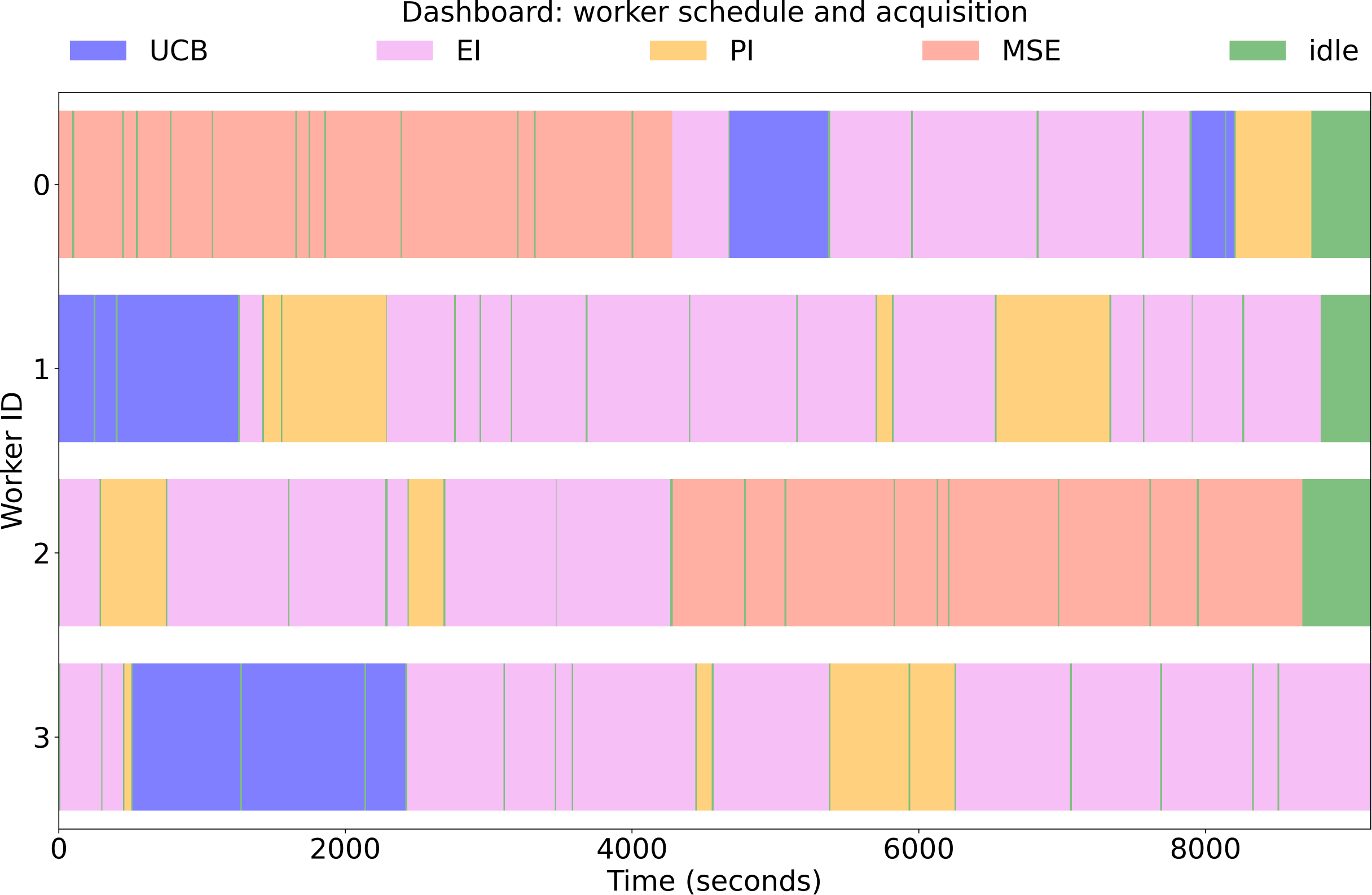}}
\caption{Comparison between batch parallel BO~\cite{tran2019pbo} and asynchronous parallel BO~\cite{tran2022aphbo} showing a substantial improvement in wallclock time reduction. By aggressively targeting the idle time for concurrent workers on high-performance computers, the wallclock time can reduce significantly. The actual improvement in wallclock time depends mainly on the stochasticity in real runtime of the application of interest.}
\label{fig:compareParallelism}
\end{figure}

We adopt the GP-Hedge scheme~\cite{hoffman2011portfolio} to sample a portfolio of acquisition functions according to their rewards. 
The main idea of GP-Hedge is to adaptively select a suitable acquisition function, whether it is EI, PI, UCB, or something else, depending their overall performance with respect to the optimization objective. 
Figure~\ref{fig:compareParallelism} compares between the batch-sequential parallelism versus the asynchronous parallelism. 
It can be observed that if the randomness of the application runtime is significant, employing an asynchronous parallel approach may lead to a substantial reduction in wallclock time, assuming there is sufficient computational resource.

\section{Crystal plasticity finite element with phenomenological constitutive model}
\label{sec:CPFEM}

A phenomenological crystal plasticity constitutive model used for face-centered cubic (FCC) crystals was first proposed by Hutchinson~\cite{hutchinson1976bounds} and extended for deformation twinning by Kalidindi~\cite{kalidindi1998incorporation}.
The plastic component is parameterized in terms of resistance $\xi$ on $N_{\text{s}}$ slip and $N_{\text{tw}}$ twin systems.
The resistances on $\alpha=1,\dots,N_{\text{s}}$ slip systems evolve from $\xi_0$ to a system-dependent saturation value and depend on shear on slip and twin systems according to
\begin{equation}
\begin{array}{lll}
\dot{\xi}^{\alpha} &=& h_0^{\text{s-s}} \left[ 1 + c_1 \left(f^{\text{tot}}_{\text{tw}}\right)^{c_2} \right] (1 + h^{\alpha}_{\text{int}}) \left[ \sum_{\alpha' = 1}^{N_{\text{s}}} \lpipe \dot{\gamma}^{\alpha'} \lpipe \ \tlpipe 1 - \frac{\xi^{\alpha'}}{\xi^{\alpha'}_{\infty}} \tlpipe^a \text{sgn}\left( 1 - \frac{\xi^{\alpha'}}{\xi^{\alpha'}_{\infty}} \right) h^{\alpha \alpha'} \right] \\
&+& \sum_{\beta'=1}^{N_{\text{tw}}} \dot{\gamma}^{\beta'} h^{\alpha \beta'} ,
\end{array}
\end{equation}
where
$f^{\text{tot}}_{\text{tw}}$ is the total twin volume fraction,
$h$ denotes the components of the slip-slip and slip-twin interaction matrices,
$h_0^{\text{s-s}}$, $h_{\text{int}}$, $c_1$, $c_2$ are model-specific fitting parameters and $\xi_{\infty}$ represents the saturated resistance.

The resistances on the $\beta = 1, \dots, N_{\text{tw}}$ twin systems evolve in a similar way,
\begin{equation}
\dot{\xi}^{\beta} = h_0^{\text{tw-s}} \left( \sum_{\alpha=1}^{N_{\text{s}}} \lpipe\gamma_{\alpha}\lpipe \right)^{c_3} \left( \sum_{\alpha'=1}^{N_{\text{s}}} \lpipe \dot{\gamma}^{\alpha'}\lpipe h^{\beta \alpha'} \right)
+ h_0^{\text{tw-tw}} \left(f^{\text{tot}}_{\text{tw}} \right)^{c_4} \left( \sum_{\beta'=1}^{N_{\text{tw}}} \dot{\gamma}^{\beta'} h^{\beta \beta'} \right),
\end{equation}
where $h_0^{\text{tw-s}}$, $h_0^{\text{tw-tw}}$, $c_3$, and $c_4$ are model-specific fitting parameters.
Shear on each slip system evolves at a rate of
\begin{equation}
\dot{\gamma}^{\alpha} = (1 - f^{\text{tot}}_{\text{tw}}) \dot{\gamma_0}^{\alpha} \tlpipe \frac{\tau^{\alpha}}{\xi^{\alpha}} \tlpipe^n \text{sgn}(\tau^{\alpha}).
\end{equation}
where slip due to mechanical twinning accounting for the unidirectional character of twin formation is computed slightly differently,
\begin{equation}
\dot{\gamma} = (1 - f^{\text{tot}}_{\text{tw}}) \dot{\gamma_0} \tlpipe \frac{\tau}{\xi} \tlpipe^n \mathcal{H}(\tau),
\end{equation}
where $\mathcal{H}$ is the Heaviside step function.
The total twin volume is calculated as
\begin{equation}
f^{\text{tot}}_{\text{tw}} = \max\left(1.0, \sum_{\beta=1}^{N_{\text{tw}}} \frac{\gamma^{\beta}}{\gamma^{\beta}_{\text{char}}} \right),
\end{equation}
where $\gamma_{\text{char}}$ is the characteristic shear due to mechanical twinning and depends on the twin system. 
Interested readers are referred to Section 6 in ~\cite{roters2019damask} for a comprehensive context of constitutive models in CPFEM. 

\section{Materials constitutive model calibration workflow}
\label{sec:ModelCalibrationWorkflow}

\subsection{A seamless integration for optimizer and forward simulations}

\begin{figure}[!htbp]
\centering
\input{fig/ConstvModelCalib_workflow}
\caption{
\textcolor{black}{An asynchronous parallel model calibration workflow for CPFEM. The workflow is mainly controlled by an advanced Bayesian optimization algorithm, called \texttt{aphBO-2GP-3BO}, to schedule an iteration on a high-performance computer. Each iteration is composed of a parameterized set of constitutive model parameters, obtained from the Bayesian optimizer. The parameterized input is then parsed into DAMASK input files. DREAM.3D is invoked to generate an ensemble of microstructure RVEs. With the complete inputs, the average loss function is evaluated and returned to the Bayesian optimizer. Iterations are scheduled asynchronously so that multiple iterations can run concurrently on a supercomputing platform to minimize the wall-clock waiting time.}}
\label{fig:ConstvModelCalib_workflow}
\end{figure}
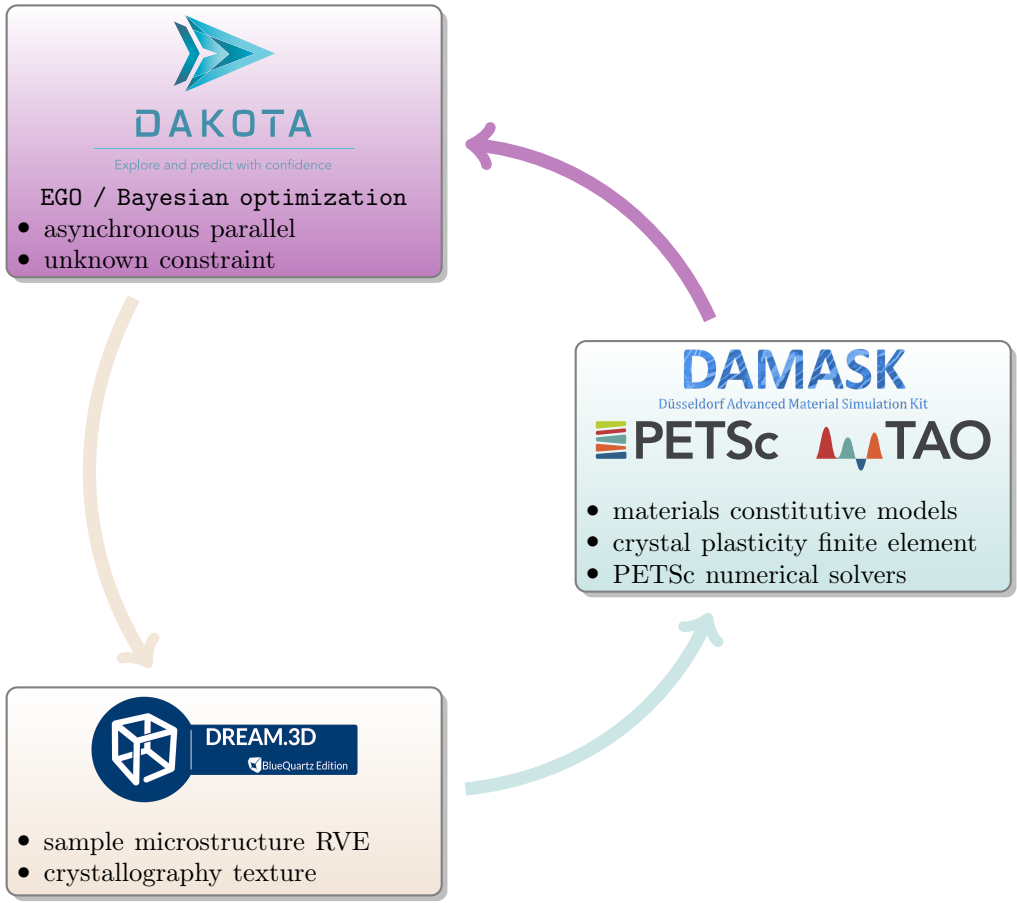

\textcolor{black}{Figure~\ref{fig:ConstvModelCalib_workflow} shows a seamless integration workflow for coupling the asynchronous parallel Bayesian optimization algorithm with an integrated forward ICME modules, composed of DREAM.3D~\cite{groeber2014dream} and DAMASK~\cite{roters2019damask} with PETSc~\cite{balay2019petsc} as the spectral solver~\cite{eisenlohr2013spectral}. The integration between DREAM.3D and DAMASK has been demonstrated in the past by Diehl et al.~\cite{diehl2017identifying}. A similar variant of \texttt{aphBO-2GP-3BO} is implemented and publicly available in DAKOTA~\cite{dalbey2022dakota}.
}

\subsection{Microstructure-ensemble average loss function}

In this paper, the loss function, to be minimized, is modeled as 
\begin{equation}
\mathcal{L}_{\text{loss}} = \frac{ \left[ \int \left(\sigma_{\text{comp}}(\varepsilon) - \sigma_{\text{exp}}(\varepsilon)\right)^2 d\varepsilon  \right]^{1/2} }{\min(\varepsilon_{\text{comp}}, \varepsilon_{\text{exp}})},
\end{equation}
where $\varepsilon_{\text{comp}}$ and $\sigma_{\text{comp}}$ are computational results from CPFEM simulation, respectively, and $\varepsilon_{\text{exp}}$ and $\sigma_{\text{exp}}$ are experimental results from experimental data, respectively.
\textcolor{black}{The computational stress used is from the Cauchy stress tensor, and the computational strain used is the true logarithmic strain.}

We measure the loss function in $L_2$ norm, normalized by the observable minimum strain $\varepsilon$ \textcolor{black}{to promote fitting over a wide range of strain $\varepsilon$. }
Due to the relatively high failure strain, CPFEM \textcolor{black}{occasionally} would fail before reaching the ultimate yield strain and therefore, the computational stress-strain curve is only compared on a much shorter computational strain interval compared to the experimental strain interval. 
To ensure that a fair comparison between early terminated CPFEM simulations and normal CPFEM simulations, we penalize the loss by dividing the normal $L_2$ loss by the $\min(\varepsilon_{\text{comp}}, \varepsilon_{\text{exp}})$, which is always less than or equal to $\varepsilon_{\text{exp}}$. 
We interpolate the numerical and experimental stress-strain equivalent curve using a cubic spline to accurately approximate the $L_2$ loss function. 
Gaussian filter \textcolor{black}{is sometimes applied} to remove noises from the experimental data. 
\textcolor{black}{
Eisenlohr et al.~\cite{eisenlohr2013spectral} and Shanthraj et al.\cite{shanthraj2019spectral} demonstrated that using the spectral solver, it is possible to achieve a fairly accurate homogenized stress-strain curve with relatively low-resolution mesh. 
This observation subsequently justifies the usage of low-resolution mesh in our approach. 
The Mat{\'e}rn-1/2 kernel, which corresponds to the exponential kernel (and the Brownian motion -- ~\cite[pp 85--86]{rasmussen2006gaussian}), is used in GP to model and overfit the loss function. 
}


\section{Case study 1: Stainless steel 304L}
\label{sec:SS304L}

\subsection{Design of numerical experiment}

The average equiaxed grain size is set as 8.0$\mu$m, following Wang et al.~\cite{wang2021study} with random crystallographic texture due to the lack of experimental data~\cite{hamza2019texture}. 
Elastic constants are adopted from Lu et al.\cite{lu2014simulation} and tabulated in Table~\ref{tab:elasticSS304L}; face-centered cubic (fcc) system is assumed.
Using DREAM.3D, a RVE of $80\mu \text{m} \times 80\mu \text{m} \times 80\mu \text{m}$ is constructed and down-sampled to $8\times8\times8$ mesh. 
A uniaxial loading condition is imposed on the RVE with $\dot{F}_{11} = 10^{-3} {\text{s}}$. 
\textcolor{black}{
At every time, concurrently, 10 CPFEM simulations are performed, under the batch settings of (6,4,0) in the \texttt{aphBO-2GP-3B} asynchronous parallel BO algorithm~\cite{tran2022aphbo}, with a batch size of 6 dedicated to the acquisition batch and a batch size of 4 dedicated to the exploration batch. 
}
The 5-dimensional input for each run $\mathbf{x}$ is the set of phenomenological constitutive model parameters $(\tau_0, \tau_\infty, h_0, n, a)$, \textcolor{black}{with the lower bounds of [1 MPa, 100 MPa, 100 MPa, 1.2, 1] and the upper bounds of [150 MPa, 12,000 MPa, 10,000 MPa, 150, 25], respectively. The reference shear rate $\dot{\gamma}_0$ is fixed at 0.001 s$^{-1}$}. 

\begin{table}[!htbp]
\centering
\caption{Elastic constants for SS304L~\cite{wang2021study}.}
\label{tab:elasticSS304L}
\begin{tabular}{lll} \hline
Elastic constants & Units  & Value  \\ \hline
$C_{11}$         & GPa   & 262.2  \\
$C_{12}$         & GPa   & 112.0  \\
$C_{44}$         & GPa   &  74.6  \\ \hline
\end{tabular}
\end{table}

\subsection{Comparison between fitted constitutive model and experimental data}

\begin{table}[!htbp]
\centering
\caption{Fitted phenomenological constitutive model parameters for SS304L.}
\label{tab:parameterSS304L}
\begin{tabular}{llll} \hline
Variable       & Description                    & Units     &  Value        \\ \hline
$\dot{\gamma}_0$   & reference shear rate       &  s$^{-1}$ &  0.001        \\
$\tau_0$       & slip resistance                &   MPa     &  90           \\
$\tau_\infty$    & saturation stress            &   MPa     &  7295.1754      \\
$h_0$        & slip hardening parameter         &   MPa     &  392.9772     \\
$n$        & strain rate sensitivity parameter  &   --      &  120          \\
$a$        & slip hardening parameter           &   --      &  8.0          \\ \hline
\end{tabular}
\end{table}

\begin{figure}[!htbp]
\centering
\includegraphics[width=\textwidth]{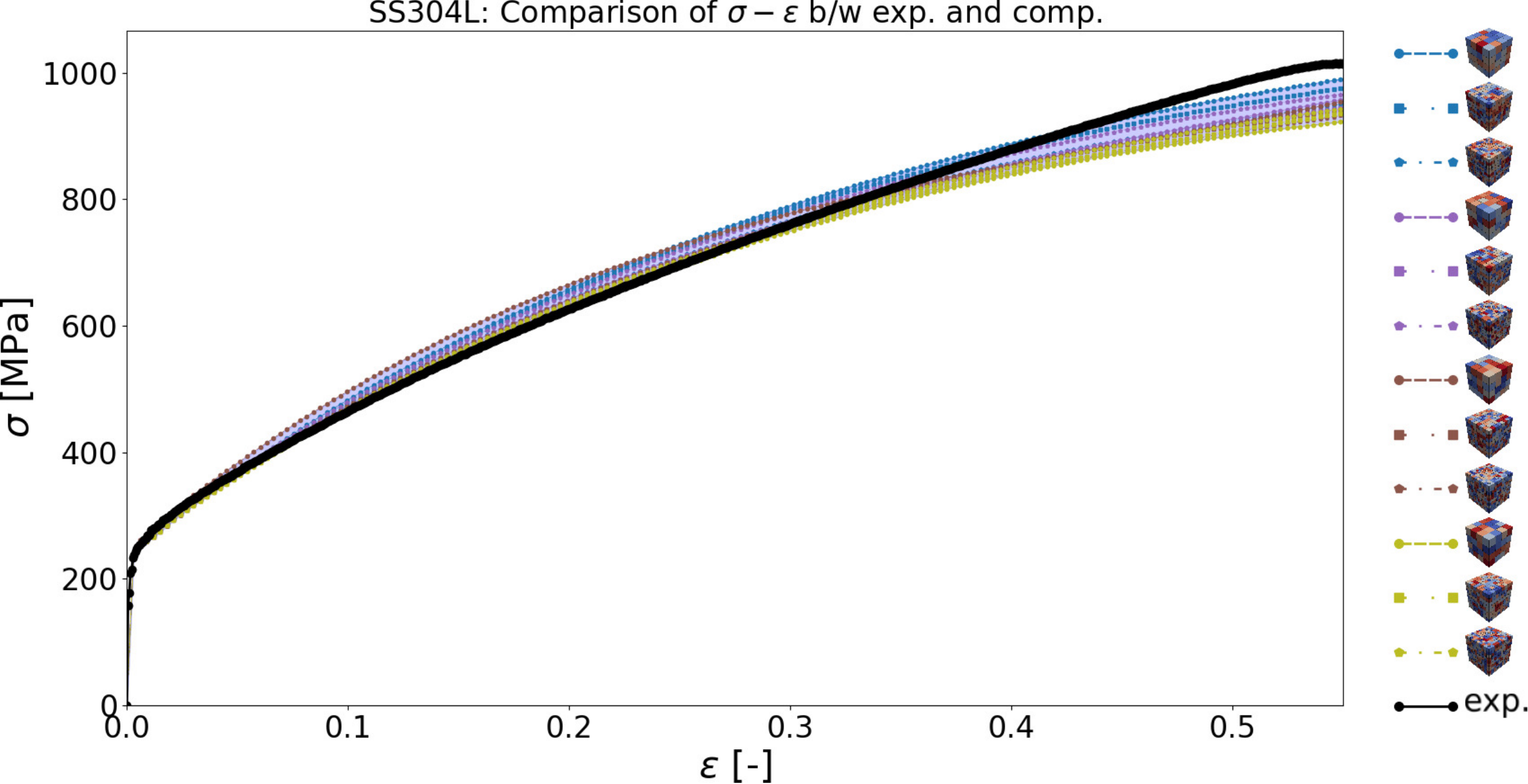}
\caption{Comparison of equivalent stress-strain curve between optimal numerical results and experimental data for SS304L, where the constitutive model parameters used are tabulated in Table~\ref{tab:parameterSS304L}. $\textcolor{black}{\medblackcircle}$ denotes experimental data. 
$\sigma/\varepsilon$ associated with RVE 1 is shown as $\textcolor{blue}{\medblackcircle}$. 
$\sigma/\varepsilon$ associated with RVE 2 is shown as $\textcolor{purple}{\medblackcircle}$. 
$\sigma/\varepsilon$ associated with RVE 3 is shown as $\textcolor{brown}{\medblackcircle}$. 
$\sigma/\varepsilon$ associated with RVE 4 is shown as $\textcolor{olive}{\medblackcircle}$. 
The material variability is shown as the \protect\tikz \protect\node [rectangle,draw, fill = blue!20] at (2.5,-2) {}; shaded region. 
Readers are referred to the color version online.}
\label{fig:compareExpCompSS304L}
\end{figure}

Table~\ref{tab:parameterSS304L} lists the optimal phenomenological constitutive model parameters used for SS304L in this study, which is found after 352 iterations. 
\textcolor{black}{
Figure~\ref{fig:compareExpCompSS304L} presents the comparison between the CPFEM numerical results with optimized constitutive model parameters and the experimental data, where the aleatory uncertainty in RVE instantiation and epistemic uncertainty in mesh resolution are both quantified. 
The \protect\tikz \protect\node [rectangle,draw, fill = blue!20] at (2.5,-2) {}; blue shaded region captures the uncertainty due to both aleatory and epistemic uncertainty in the optimization process. 
An ensemble of 4 RVEs are also shown in Figure~\ref{fig:compareExpCompSS304L} as $\textcolor{blue}{\medblackcircle}$, $\textcolor{purple}{\medblackcircle}$, $\textcolor{brown}{\medblackcircle}$, $\textcolor{olive}{\medblackcircle}$, to highlight the natural variability and inherent randomness of microstructures on the $\sigma/\varepsilon$ curves. 
Overall, the phenomenological constitutive model without twinning fits well in the high-strain $(\varepsilon > 0.10)$ plasticity regime but does not accurately capture the hardening behaviors. The low-strain $(\varepsilon < 0.10)$ plasticity regime is poorly captured, due to the lack of intricacy in the constitutive model. 
}

\section{Case study 2: Tantalum phenomenological constitutive model}
\label{sec:Tantalum}

\subsection{Design of numerical experiment}

\textcolor{black}{
In this case study, we are interested in fitting the bcc Tantalum, with the average grain size of 50$\mu$m. 
Elastic constants are adopted from~\cite{lim2014grain,duesbery1998plastic} and tabulated in Table~\ref{tab:elasticTantalum}; body-centered cubic (bcc) system is assumed.
Using DREAM.3D, a RVE of $80\mu \text{m} \times 80\mu \text{m} \times 80\mu \text{m}$ is constructed and down-sampled to $8\times8\times8$ mesh. 
A uniaxial loading condition is imposed on the RVE with $\dot{F}_{11} = 10^{-3} {\text{s}}$. 
No twinning is considered for quasi-static loading at the ambient temperature. Even though not considered in this study of calibrating phenomenological constitutive models, calculating Peierls stresses in tantalum have been performed in the past through molecular dynamics simulation~\cite{wang2004calculating,anglade2005computation}. The 5-dimensional input for each run $\mathbf{x}$ is the set of phenomenological constitutive model parameters $(\tau_0, \tau_\infty, h_0, n, a)$, with the lower bounds of [1 MPa, 1 MPa, 1 MPa, 1.2, 1] and the upper bounds of [1,000 MPa, 10,000 MPa, 10,000 MPa, 150, 200], respectively. The reference shear rate $\dot{\gamma}_0$ is fixed at 0.001 s$^{-1}$. 
At every time, concurrently, 24 CPFEM simulations are performed, under the batch settings of (10,10,4) in the \texttt{aphBO-2GP-3B} asynchronous parallel Bayesian optimization algorithm~\cite{tran2022aphbo}, with a batch size of 10 dedicated to the acquisition batch, a batch size of 4 dedicated to the objective exploration batch, and a batch size of 4 dedicated to the classification exploration batch. 
}

\begin{table}[!htbp]
\centering
\caption{Elastic constants for Tantalum~\cite{lim2014grain,duesbery1998plastic}.}
\label{tab:elasticTantalum}
\begin{tabular}{lll} \hline
Elastic constants & Units  & Value  \\ \hline
$C_{11}$         & GPa   & 267    \\
$C_{12}$         & GPa   & 161    \\
$C_{44}$         & GPa   & 82.5   \\ \hline
\end{tabular}
\end{table}

\subsection{Comparison between fitted constitutive model and experimental data}

\begin{table}[!htbp]
\centering
\caption{\textcolor{black}{Fitted phenomenological constitutive model parameters for Tantalum.}}
\label{tab:parameterTantalum}
\begin{tabular}{llll} \hline
Variable       & Description                    & Units     &  Value        \\ \hline
$\dot{\gamma}_0$   & reference shear rate       &  s$^{-1}$ &  0.001        \\
$\tau_0$       & slip resistance                &   MPa     &  67.4641           \\
$\tau_\infty$    & saturation stress            &   MPa     &  7295.1754      \\
$h_0$        & slip hardening parameter         &   MPa     &  1959.1320     \\
$n$        & strain rate sensitivity parameter  &   --      &  45.2726          \\
$a$        & slip hardening parameter           &   --      &  200.0          \\ \hline
\end{tabular}
\end{table}

\begin{figure}[!htbp]
\centering
\includegraphics[width=\textwidth]{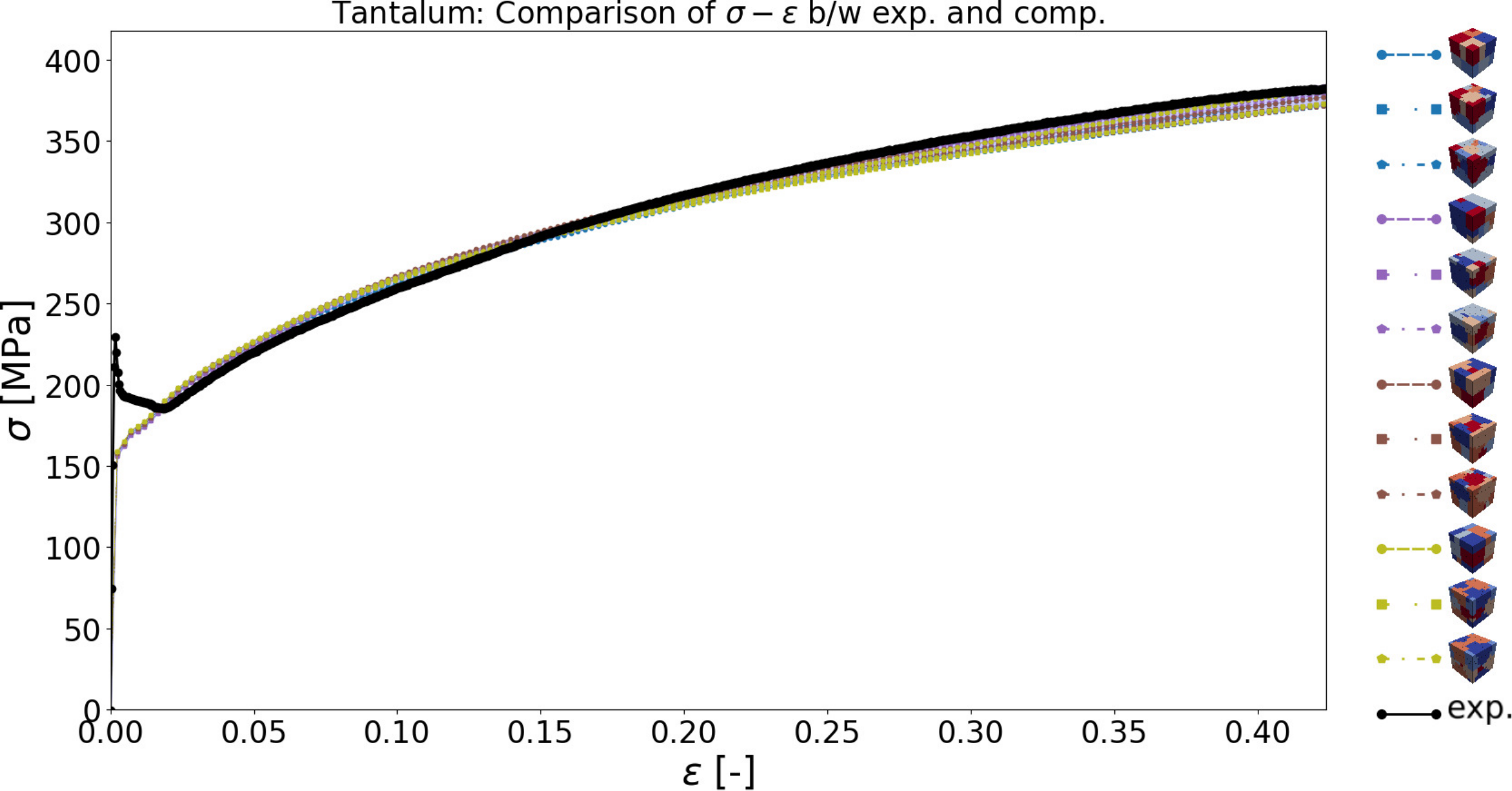}
\caption{
\textcolor{black}{
Comparison of equivalent stress-strain curve between optimal numerical results and experimental data for Tantalum, where the constitutive model parameters used are tabulated in Table~\ref{tab:parameterTantalum}. $\textcolor{black}{\medblackcircle}$ denotes experimental data. 
$\sigma/\varepsilon$ associated with RVE 1 is shown as $\textcolor{blue}{\medblackcircle}$. 
$\sigma/\varepsilon$ associated with RVE 2 is shown as $\textcolor{purple}{\medblackcircle}$. 
$\sigma/\varepsilon$ associated with RVE 3 is shown as $\textcolor{brown}{\medblackcircle}$. 
$\sigma/\varepsilon$ associated with RVE 4 is shown as $\textcolor{olive}{\medblackcircle}$. 
The material variability is shown as the \protect\tikz \protect\node [rectangle,draw, fill = blue!20] at (2.5,-2) {}; shaded region. 
Readers are referred to the color version online.}
}
\label{fig:compareExpCompTantalum}
\end{figure}

\textcolor{black}{
Table~\ref{tab:parameterTantalum} lists the optimal phenomenological constitutive model parameters used for Tantalum in this study, which is found after 1135 iterations. 
Figure~\ref{fig:compareExpCompTantalum} presents the comparison between the CPFEM numerical results with optimized constitutive model parameters and the experimental data, where the aleatory uncertainty in RVE instantiation and epistemic uncertainty in mesh resolution are both quantified. 
The \protect\tikz \protect\node [rectangle,draw, fill = blue!20] at (2.5,-2) {}; blue shaded region captures the uncertainty due to both aleatory and epistemic uncertainty in the optimization process. 
An ensemble of 4 RVEs are also shown in Figure~\ref{fig:compareExpCompTantalum} as $\textcolor{blue}{\medblackcircle}$, $\textcolor{purple}{\medblackcircle}$, $\textcolor{brown}{\medblackcircle}$, $\textcolor{olive}{\medblackcircle}$, to highlight the natural variability and inherent randomness of microstructures on the $\sigma/\varepsilon$ curves. 
Overall, the numerical results agree relatively well with the experimental data, given a long interval of strain, except for the immediate drop in stress right after the elastic deformation. 
It should be noted that the experimental data are filtered to exclude the peak and this immediate sharp drop; in other words, we exclude the data between the strain $\varepsilon = [9.3667 \cdot 10^{-4}, 1.8572 \cdot 10^{-2}]$.
}

\section{Case study 3: Cantor CrMnFeCoNi high-entropy alloy phenomenological constitutive model}
\label{sec:CantorHEA}

\subsection{Design of numerical experiment}

\begin{table}[!htbp]
\centering
\caption{Elastic constants for Cantor alloy~\cite{gludovatz2015processing,laplanche2018elastic}.}
\label{tab:elasticCantorAlloy}
\begin{tabular}{lll} \hline
Elastic constants & Units  & Value  \\ \hline
$C_{11}$         & GPa   & 172  \\
$C_{12}$         & GPa   & 108  \\
$C_{44}$         & GPa   & 92   \\ \hline
\end{tabular}
\end{table}

\textcolor{black}{
In this case study, we are interested in fitting a simple phenomenological constitutive model without twinning for the Cantor alloy, which is also known as the CrMnFeCoNi high-entropy alloy. 
The experimental data is obtained from Chen et al~\cite{chen2020real}. 
We follow ~\cite{gludovatz2015processing,laplanche2018elastic} for elastic constants, which are tabulated in Table~\ref{tab:elasticCantorAlloy}. Only fcc system is considered in this case study, even though fcc, hcp (hexagonal closed packed)~\cite{chen2020real}, and bcc~\cite{zeng2021mechanical} phases are observed experimentally. 
Using DREAM.3D, a RVE of $80\mu \text{m} \times 80\mu \text{m} \times 80\mu \text{m}$ is constructed and down-sampled to $8\times8\times8$ mesh. 
A uniaxial loading condition is imposed on the RVE with $\dot{F}_{11} = 10^{-3} {\text{s}}$. 
Average grain size of 7$\mu$m from literature~\cite{thurston2017effect,chen2019grain,rackwitz2020effects}. 
No twinning is considered in fitting the phenomenological constitutive model. 
At every time, concurrently, 24 CPFEM simulations are performed, under the batch settings of (10,10,0) in the \texttt{aphBO-2GP-3B} asynchronous parallel Bayesian optimization algorithm~\cite{tran2022aphbo}, with a batch size of 10 dedicated to the acquisition batch, a batch size of 10 dedicated to the objective exploration batch. 
The 5-dimensional input for each run $\mathbf{x}$ is the set of phenomenological constitutive model parameters $(\tau_0, \tau_\infty, h_0, n, a)$. 
}

\subsection{Comparison between fitted constitutive model and experimental data}

\begin{table}[!htbp]
\centering
\caption{\textcolor{black}{Fitted phenomenological constitutive model parameters for Cantor alloy (without twinning).}}
\label{tab:parameterCantorAlloy}
\begin{tabular}{llll} \hline
Variable       & Description                    & Units     &  Value         \\ \hline
$\dot{\gamma}_0$   & reference shear rate       &  s$^{-1}$ &  0.001         \\
$\tau_0$       & slip resistance                &   MPa     &  1.0           \\
$\tau_\infty$    & saturation stress            &   MPa     &  10000.0       \\
$h_0$        & slip hardening parameter         &   MPa     &  1959.1320     \\
$n$        & strain rate sensitivity parameter  &   --      &  41.2673       \\
$a$        & slip hardening parameter           &   --      &  128.3608      \\ \hline
\end{tabular}
\end{table}

\begin{figure}[!htbp]
\centering
\includegraphics[width=\textwidth]{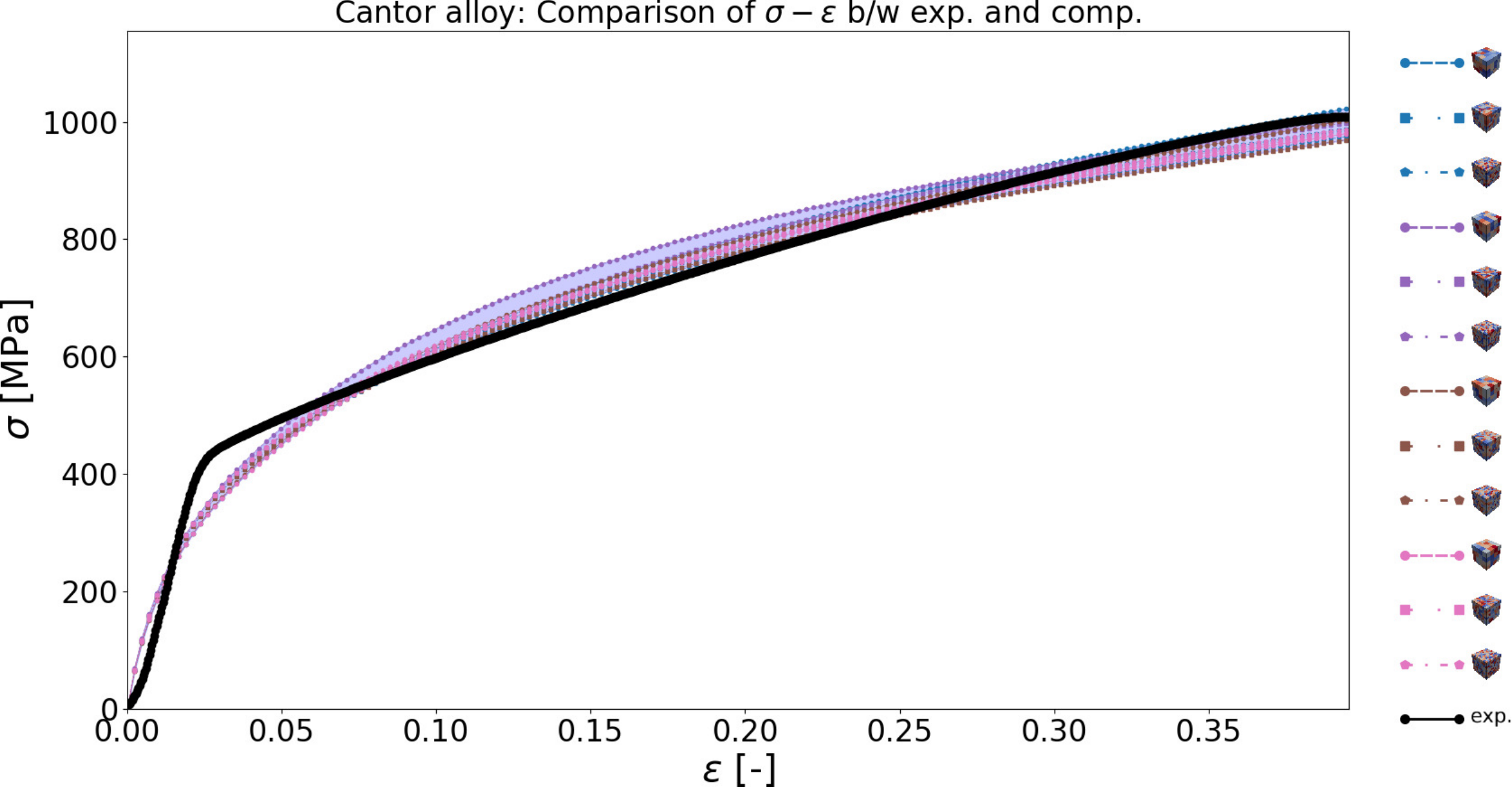}
\caption{
\textcolor{black}{
Comparison of equivalent stress-strain curve between optimal numerical results and experimental data for Cantor alloy, where the constitutive model parameters used are tabulated in Table~\ref{tab:parameterCantorAlloy}. $\textcolor{black}{\medblackcircle}$ denotes experimental data~\cite{chen2020real}. 
$\sigma/\varepsilon$ associated with RVE 1 is shown as $\textcolor{blue}{\medblackcircle}$. 
$\sigma/\varepsilon$ associated with RVE 2 is shown as $\textcolor{purple}{\medblackcircle}$. 
$\sigma/\varepsilon$ associated with RVE 3 is shown as $\textcolor{brown}{\medblackcircle}$. 
$\sigma/\varepsilon$ associated with RVE 4 is shown as $\textcolor{olive}{\medblackcircle}$. 
The material variability is shown as the \protect\tikz \protect\node [rectangle,draw, fill = blue!20] at (2.5,-2) {}; shaded region. 
Readers are referred to the color version online.}
}
\label{fig:compareExpCompCantorAlloy}
\end{figure}

\textcolor{black}{
Table~\ref{tab:parameterCantorAlloy} lists the optimal phenomenological constitutive model parameters used for the Cantor alloy in this study, which is found after 969 iterations. 
Figure~\ref{fig:compareExpCompCantorAlloy} presents the comparison between the CPFEM numerical results with optimized constitutive model parameters and the experimental data, where the aleatory uncertainty in RVE instantiation and epistemic uncertainty in mesh resolution are both quantified. 
The \protect\tikz \protect\node [rectangle,draw, fill = blue!20] at (2.5,-2) {}; blue shaded region captures the uncertainty due to both aleatory and epistemic uncertainty in the optimization process. 
An ensemble of 4 RVEs are also shown in Figure~\ref{fig:compareExpCompCantorAlloy} as $\textcolor{blue}{\medblackcircle}$, $\textcolor{purple}{\medblackcircle}$, $\textcolor{brown}{\medblackcircle}$, $\textcolor{olive}{\medblackcircle}$, to highlight the natural variability and inherent randomness of microstructures on the $\sigma/\varepsilon$ curves. 
Uncertainty quantification on constitutive models in CPFEM has been carried out in the past~\cite{tran2022microstructure}. 
However, it seems that the phenomenological constitutive model without twinning in this case study is inadequate to model the complex materials behaviors of the Cantor alloy, which suggests a more complicated dislocation-density-based constitutive, e.g.~\cite{steinmetz2013revealing}, may better fit its behaviors over a wide range of compositions and temperature. 
}




\section{Discussion \& Conclusion}
\label{sec:DiscussionConclusion}

In this paper, we apply an asynchronous parallel constrained BO algorithm~\cite{tran2022aphbo} to calibrate phenomenological constitutive model \textcolor{black}{for fcc SS304L, bcc Tantalum, and fcc Cantor alloy.} 
\textcolor{black}{To maximize the acquisition function, we rely on the CMA-ES algorithms~\cite{hansen2001completely,hansen2003reducing,hansen2004evaluating}, which is also gradient-free to search for the next most informative input parameters.} 
Compared to other sequential optimization algorithms used in the literature, the \texttt{aphBO-2GP-3B} algorithm exploits the computational resource by \textit{two} layers of parallelisms: the \textit{optimization parallelism} and the \textit{message passing interface (MPI) parallelism}, where the former takes a priority over the later due to diminishing return by Amdahl's law~\cite{hill2008amdahl}. 

\textcolor{black}{
Model calibration methods are mainly divided into two classes: deterministic and statistical~\cite{dalbey2022dakota}. On one hand, the deterministic model calibration seeks a unique set of parameters that optimally matches experimental data, where optimality is typically measured in $L_2$, as done in this paper. On the other hand, the statistical model calibration, such as Bayesian inference~\cite{zhang2019numerical}, seeks a statistical characterization of parameters that are most consistent with the data.
}
\textcolor{black}{
In this study, the only constraint is unknown, where the CPFEM simulation may not return a numerical value, e.g. \texttt{NaN}. This behavior occasionally happens, depending on the user-defined optimization bounds. This type of constraints (i.e. blind constraints) is conveniently handled thanks to the internal binary classifier that penalize the acquisition function~\cite{tran2022aphbo} to navigate the complex acquisition function map.
}

\textcolor{black}{Multi-fidelity GP~\cite{kennedy2000predicting,kennedy2001bayesian} could also be used to combine the computational and experimental data, either at the same length-scale or in multi-scale manner~\cite{tran2020multi}. In this study, we fix the mesh size due to its insensitivity, based on prior DAMASK studies, e.g.~\cite{eisenlohr2013spectral}. However, generalizing to an asynchronous parallel constrained multi-fidelity GP/BO framework with adaptive coarse and fine meshes is also possible and makes more sense, thus remaining as an interesting future research direction. 
}
\textcolor{black}{In this study, the only constraint is unknown, where the CPFEM simulation may not return a numerical value, e.g. \texttt{NaN}. This behavior occasionally happens, depending on the user-defined optimization bounds. This type of constraints (i.e. blind constraints) is conveniently handled thanks to the internal binary classifier that penalize the acquisition function~\cite{tran2022aphbo} to navigate the complex acquisition function map.} 

Based on the optimized \textcolor{black}{constitutive model} parameters, we conduct a simple \textcolor{black}{verification and validation} study to investigate the effects of aleatory uncertainty from random RVEs and epistemic uncertainty from mesh-resolution. 
The homogenized \textcolor{black}{computational} results are \textcolor{black}{shown to be reproducibly consistent} with the experimental data. 
\textcolor{black}{Stainless steel 304L and Tantalum results show an excellent fit without twinning; however, Cantor alloy results shows a relatively poor fit due to an overly simplified constitutive model for a complex materials system.} 

We conclude that the asynchronous parallel constrained BO~\cite{tran2022aphbo} is capable of calibrating phenomenological constitutive models for CPFEM, which paves way for further research in the future. 


\section*{Acknowledgment}

The views expressed in the article do not necessarily represent the views of the U.S. Department of Energy or the United States Government. Sandia National Laboratories is a multimission laboratory managed and operated by National Technology and Engineering Solutions of Sandia, LLC., a wholly owned subsidiary of Honeywell International, Inc., for the U.S. Department of Energy's National Nuclear Security Administration under contract DE-NA-0003525. 

A.T. thanks Prof. Robert O. Ritchie (University of Berkeley / Lawrence Berkeley National Laboratory) and Prof. Bernd Gludovatz (University of New South Wales) for sharing their Cantor alloys experimental data, Prof. Philip Eisenlohr (Michigan State University) and Dr. Pieterjan Robbe (Sandia National Laboratories) for numerous fruitful discussions. 

\textcolor{black}{The authors are extremely grateful to the editors and the two anonymous reviewers for their valuable comments and suggestions, which have greatly improved the quality of our manuscript.}

\bibliography{lib}


\end{document}

%% file: fig/ConstvModelCalib_workflow.tex
\begin{center}

\smartdiagramset{
set color list={violet!50, brown!20, teal!20},
circular distance=5.0cm,
font=\large,
text width=5.5cm,
module minimum width=3.5cm,
module minimum height=1.5cm,
arrow line width=5pt,
arrow tip=to,
}
\centering
\smartdiagram[circular diagram]{
  \includegraphics[height=60px]{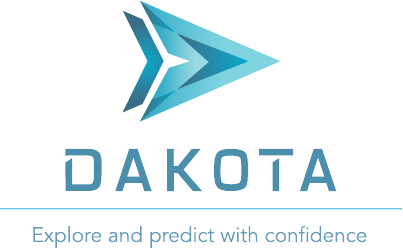} \\
  \normalsize \textbf{\texttt{EGO / Bayesian optimization}} \normalsize \\
  \begin{itemize}
    \item asynchronous parallel
    \item unknown constraint
  \normalsize
  \end{itemize}
  ,
  \includegraphics[height=40px]{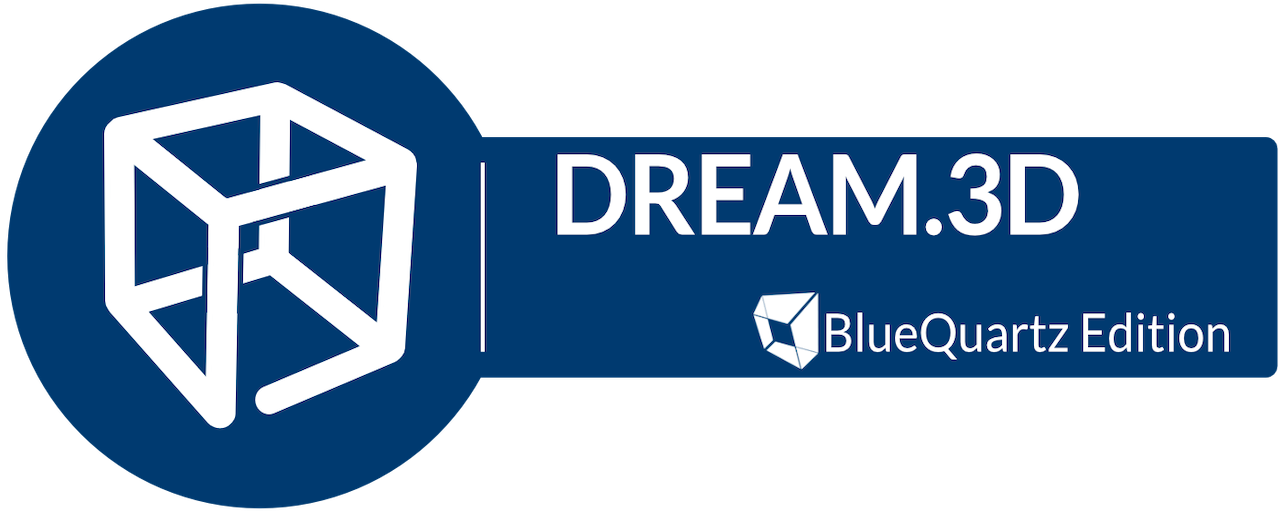} \\
  \begin{itemize}
    \item sample microstructure RVE
    \item crystallography texture
  \normalsize
  \end{itemize}
  ,
  \includegraphics[height=25px]{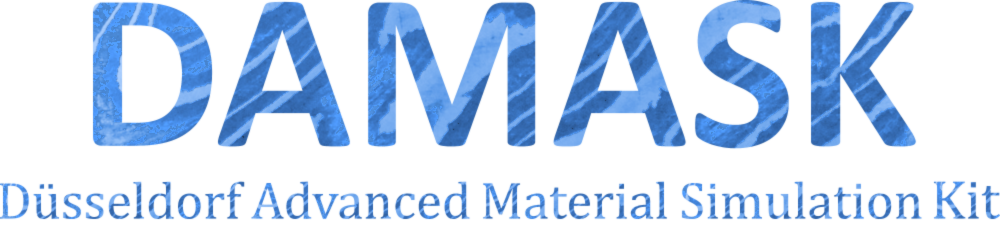} \\
  \includegraphics[height=20px]{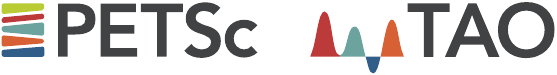} \\
  \begin{itemize}
    \item materials constitutive models
    \item crystal plasticity finite element
    \item PETSc numerical solvers
  \normalsize
  \end{itemize}
}


\end{center}